\documentclass[aps,prl,superscriptaddress,twocolumn,10pt]{revtex4-1}

\usepackage[utf8]{inputenc}
\usepackage[version=3]{mhchem}
\usepackage{amsmath}
\usepackage{amsfonts}
\usepackage{hyperref}
\usepackage{braket}
\usepackage{lipsum}
\usepackage{url}
\usepackage[draft]{graphicx}
\usepackage{lipsum}
\usepackage{color}
\usepackage{float}
\usepackage[export]{adjustbox}
\usepackage{wrapfig}
\usepackage{array,multirow}
\usepackage{tabularx}
\usepackage{bm}
\usepackage{chemfig}
\usepackage{booktabs}

\usepackage{relsize}  

\makeatletter
\renewcommand\frontmatter@abstractwidth{\dimexpr\textwidth-1in\relax}
\makeatother

\usepackage[normalem]{ulem}
\definecolor{my_red}{rgb}{.7,0,0}
\definecolor{my_green}{rgb}{0,.7,0}
\definecolor{my_blue}{rgb}{0,0,.7}

\newcommand{\CM}{\ensuremath{_\text{CM}}}

\begin{abstract}
Charge migration (CM) is a coherent attosecond process that involves the movement of localized holes across a molecule. To determine the relationship between a molecule's structure and the CM dynamics it exhibits, we perform systematic studies of {\it para}-functionalized bromobenzene molecules (\ce{X-C6H4-R}) using real-time time-dependent density functional theory. We initiate valence-electron dynamics by emulating rapid strong-field ionization leading to a localized hole on the bromine atom. The resulting CM, which takes on the order of 1~fs, occurs via an X~localized $\rightarrow$ \ce{C6H4} delocalized $\rightarrow$ R~localized mechanism. Interestingly, the hole contrast on the acceptor functional group increases with increasing electron donating strength. This trend is well-described by the Hammett sigma value of the group, which is a commonly used metric for quantifying the effect of functionalization on the chemical reactivity of benzene derivatives. These results suggest that simple attochemistry principles and a density-based picture can be used to predict and understand CM.
\end{abstract}

\begin{document}

\author{Aderonke S. Folorunso}
\affiliation{Department of Chemistry, Louisiana State University, Baton Rouge, LA 70803}

\author{Fran\c{c}ois Mauger}
\affiliation{Department of Physics and Astronomy, Louisiana State University, Baton Rouge, LA 70803}

\author{Kyle A. Hamer}
\affiliation{Department of Physics and Astronomy, Louisiana State University, Baton Rouge, LA 70803}

\author{Denawakage D Jayasinghe}
\affiliation{Department of Chemistry, Louisiana State University, Baton Rouge, LA 70803}

\author{Imam Wahyutama}
\affiliation{Department of Physics and Astronomy, Louisiana State University, Baton Rouge, LA 70803}

\author{Justin R. Ragains}
\affiliation{Department of Chemistry, Louisiana State University, Baton Rouge, LA 70803}

\author{Robert R. Jones}
\affiliation{Department of Physics, University of Virginia, Charlottesville, VA 22904}

\author{Louis F. DiMauro}
\affiliation{Department of Physics, The Ohio State University, Columbus, OH 43210}

\author{Mette B. Gaarde}
\affiliation{Department of Physics and Astronomy, Louisiana State University, Baton Rouge, LA 70803}

\author{Kenneth J. Schafer}
\affiliation{Department of Physics and Astronomy, Louisiana State University, Baton Rouge, LA 70803}

\author{Kenneth Lopata}
\email{klopata@lsu.edu}
\affiliation{Department of Chemistry, Louisiana State University, Baton Rouge, LA 70803}
\affiliation{Center for Computation and Technology, Louisiana State University, Baton Rouge, LA 70803}

\title{
Attochemistry Regulation of Charge Migration
}

\maketitle

\newpage
Charge migration, first identified in the pioneering work of Cederbaum and Zobeley~\cite{cederbaum1999ultrafast}, is the coherent movement of an electron density hole across a molecule~\cite{Robb2014glycine,calegari2014ultrafast,kraus2015measurement,golubev2015control,breidbach2005universal,kling2008attosecond,Folorunso2021Molecular}. These holes can be created using neutral excitation{\cite{calegari2015ultrafast,BNbriged2011}} or ionization via inner shell\cite{despre2015attosecond,remacle2006electronic} or strong field\cite{SFI2017,SFI2019} processes. CM, which occurs on an attosecond time scale, typically does not involve nuclear motion. These short-time dynamics are postulated to influence longer-time photochemical processes such as photosynthesis, photocatalysis, and light harvesting\cite{Eberhard2008photo,cederbaum1999ultrafast,BNbriged2011}. Furthermore, these dynamics are expected to modulate photochemical reactivity, since the distribution of charge in a molecule influences nuclear motion\cite{fRemacle1998}.  
Since its discovery, there have been numerous theoretical studies of CM in small molecules \cite{lunnemann2008ultrafast,kuleff2013electron,remacle2006electronic,lara2017role,calegari2015ultrafast}, along with some experimental studies using high harmonic generation and pump-probe ionization methods \cite{kraus2015measurement,calegari2014ultrafast,fmartin2018}. Organic aromatic molecules are especially promising, since they support facile CM due to their conjugated $\pi$-electron system~\cite{bruner2017attosecond,despre2015attosecond,despre2021}, within which the hole can be viewed as hopping between $\pi$-bonds\cite{Folorunso2021Molecular,BNbriged2011}. Many questions remain, however, concerning the relationship between structure/chemical functionalization and the CM dynamics these molecule can support. Previous studies have largely been on a case-by-case basis, and invoked some form of state-based picture for explanation. A chemistry-based interpretation of CM, which uses concepts such as electron donating/withdrawing to explain trends, remains relatively undeveloped.

To address this problem, we present a systematic first-principles simulation study of CM in functionalized bromobenzene derivatives, and use this to develop a set of ``attochemistry'' principles, which draw on simple chemical ideas to predict and understand CM in this family of molecules. Bromobenzene is a good prototypical CM system, as the Br atom supports the creation  of a localized hole either via strong-field~\cite{SFI2017,SFI2019} or inner-shell ionization~\cite{keller1986shape,boll2016charge,schnorr2014electron}. Additionally, benzene can be easily introduced into the gas phase, and has CM oscillations that survive more than 10~fs, despite the presence of nuclear dynamics\cite{despre2015attosecond}. Moreover, benzene is highly customizable and can be modified with a range of functional groups at the {\it ortho, meta}, and {\it para} positions to yield stable compounds, many of which are either commercially available or easily synthesized. These compounds are also the building blocks for more complicated systems, such as biomolecules and polycyclic aromatic hydrocarbons (PAHs). 
Thus, determining structure/CM relationships in the bromobenzene series helps
form a bridge between the chemical properties of a molecule and the attosecond dynamics it supports, which can be generalizable to a wide range of systems. These relationships, in turn, will be useful for guiding choice of molecules for future CM study, as well as for interpreting measurements.
\begin{figure*}
\centering
  \includegraphics[width=0.9\linewidth]{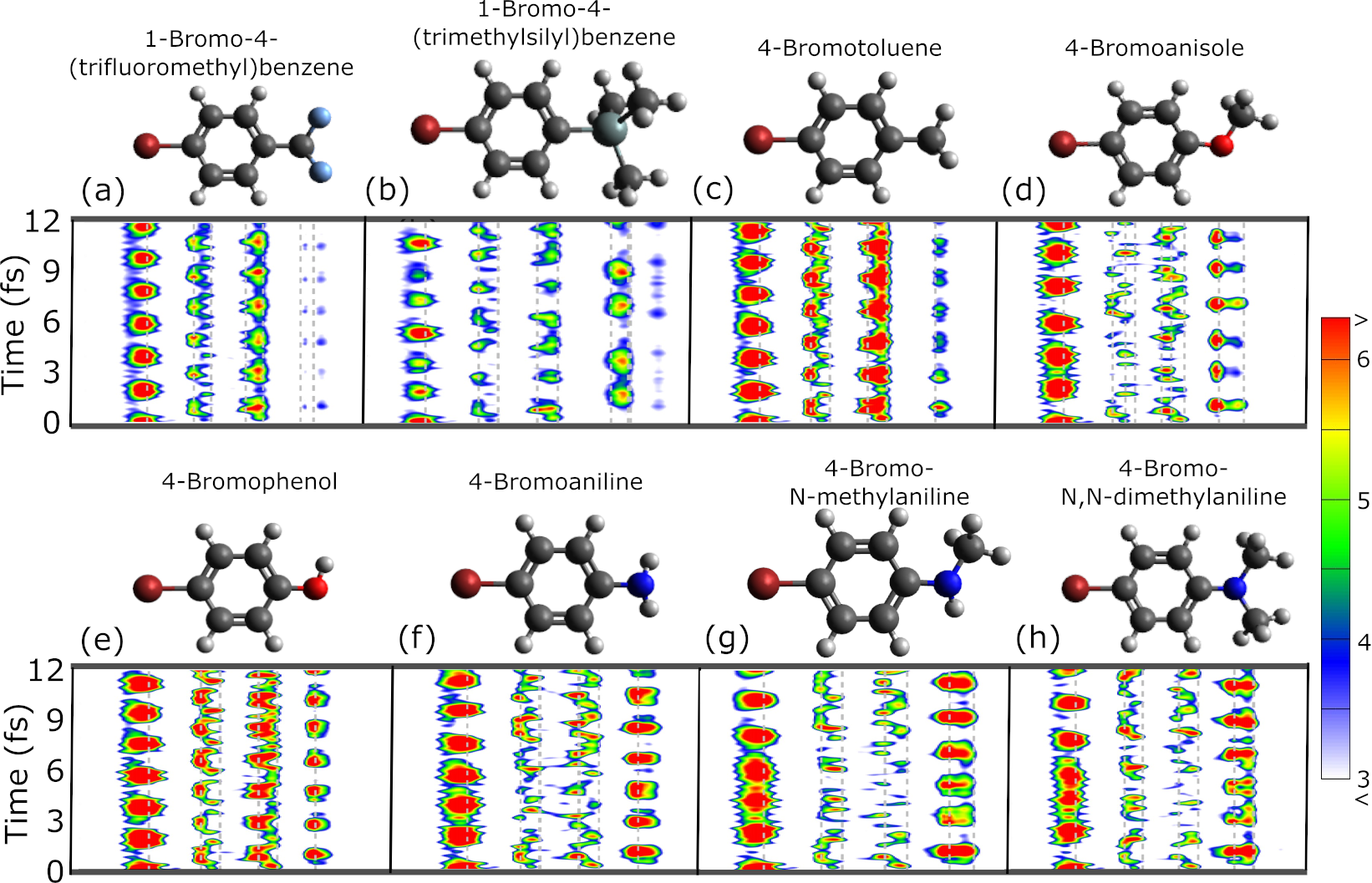}
  \caption{The effect of functional group on CM dynamics in bromobenzene derivatives. Panels (a-h) show the positive part of the time-dependent perpendicular-integrated hole densities following sudden ionization from the Br atom. As the electron donating strength of the {\it para} group increases ((a) to (h)), the hole contrast on the end group increases.}
  \label{fig:funct}
\end{figure*}

To simulate CM, we use real-time time-dependent density functional theory (RT-TDDFT) as implemented~\cite{lopata2011modeling,Yabana1996rrttddft,Isborn2007rrttddft} in NWChem~\cite{apra2020nwchem}.  For all simulations, we use the hybrid PBE0 functional~\cite{Adamo1999Pbe0}, cc-pVDZ for H/C and Stuttgart RLC ECP for Br with a time step of 0.2~a.u.\ (0.005~fs) and 1000~a.u.\ (24~fs) to propagate the dynamics. For the initial state, we use a sudden approximation for the strong-field ionization (SFI) step by creating a hole on the bromine atom at $t=0$ using constrained DFT (cDFT)~\cite{Voorhis2009cdft}. Knowing that SFI from brominated organic molecules results in a Br localized hole~\cite{SFI2019,Folorunso2021Molecular}, we use cDFT to minimize the energy with the constraint that the Br atom has a +1 charge. In practice, this ionization ``simulant'' mixes multiple orbitals to give multi-electron, multideterminant-like excitations, akin to the self-consistent field ($\Delta$SCF) method for excited states~\cite{Gill2009SCF,liang2017accurate,ramos2018low}. In a state picture, this localization process puts the molecule in an intricate superposition of ionic states which results in coherent CM dynamics. This bypasses well-known challenges when using TDDFT with adiabatic exchange-correlation functionals to drive systems far from equilibrium~\cite{fuks2013dynamics,maitra2006long,giesbertz2008failure,provorse2015peak,fuks2015time,raghunathan2012lack}. We previously explored the role of the initial hole localization on CM, which can be understood in a nonlinear dynamics framework\cite{mauger2022charge}.

To interpret the resulting dynamics, we use the hole density, $\rho^H(\bm{r},t)$, computed by subtracting the neutral ground state density from the time-dependent cation density: $\rho^{H}(\bm{r},t) = \rho^0(\bm{r})-\rho^+(\bm{r},t)$.
The hole density is then integrated over directions transverse to the CM axis (long axis of the molecule) for easier visualization and for computation of various metrics. To display a clearer time-dependence map that shows the CM modes, we remove the high-frequency contributions in all time-dependent plots of $\rho^{\rm H}(r,t)$, using filtering via convolution with a $\sin^2$ temporal window with a 0.8~fs total duration.

\begin{figure*}
  \begin{center}
    \includegraphics[width=0.75\linewidth]{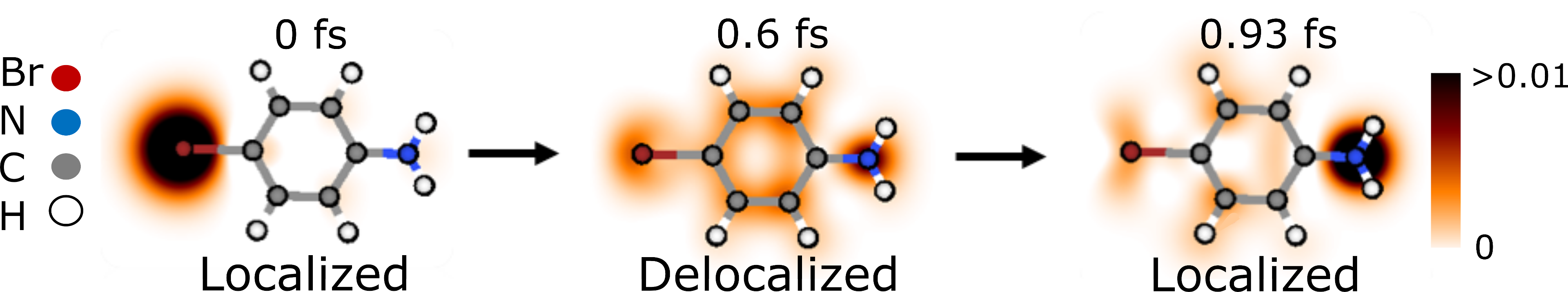}
     \caption{Snapshot of the positive part of the hole density in 4-bromoaniline at 1.0 \AA\ above the plane of the molecule immediately following ionization. The hole undergoes a localized $\rightarrow$ delocalized $\rightarrow$ localized charge migration process that takes 0.93~fs.}
     \label{fig:mecha}
  \end{center}
\end{figure*}

For the cases that do result in CM, hole density maps can be used to compute a range of physically relevant metrics. The CM time ($t\CM$) is the time it takes for the hole to travel from the Br atom to the acceptor group \ce{-R}. The CM distance and speed can be similarly defined~\cite{Folorunso2021Molecular}, but are not used in this study. To quantify the degree of hole localization on the \ce{-R} group, we use the hole contrast $\Gamma$, a dimensionless quantity that is expected to be correlated with the sensitivity of experimental probes of the density around \ce{-R}.
First, the hole density is computed by integrating the hole density 1~\AA~above the plane of the molecule, and then integrating the hole number on the acceptor \ce{-R}. This integration selects the part of the density involved in the CM, which mainly occurs in the $\pi$ system of the molecule. $\Gamma$ is then obtained by fitting to an offset oscillation with the same frequency as the CM: $ n^H_{\text{R}} = A+B \sin \left( \omega_{\text{CM}} t + \phi \right) $\cite{Folorunso2021Molecular}, where $ n^H_{\text{R}} $ is the number of holes on \ce{-R}. The hole contrast on \ce{-R} is given by ratio $\Gamma = B/A$. 

Fig.~\ref{fig:funct} shows the integrated hole density time plots for a range of functionalized bromobezene molecules. The functional groups are ordered (a)$\rightarrow$(h) by increasing electron donating strength: 1-Bromo-4-(trifluoromethyl)benzene ($\ce{BrC6H4CF3}$), 1-Bromo-4-(trimethylsilyl)benzene ($\ce{BrC6H4Si(CH3)3}$), 4-bromotoluene ($\ce{BrC6H4CH3}$), 4-bromoanisole ($\ce{BrC6H4OCH3}$),  4-bromophenol ($\ce{BrC6H4OH}$), 4-bromoaniline ($\ce{BrC6H4NH2}$), 4-Bromo-{\it N}-methylaniline ($\ce{BrC6H4NHCH3}$) and 4-Bromo-{\it N,N}-dimethylaniline ($\ce{BrC6H4N(CH3)2}$). 

We now qualitatively describe the dynamics in these systems. In Fig. \ref{fig:funct}(a), the \ce{-CF3} case is distinct in that it does not support CM.  This is a consequence of this group's strong electron withdrawing strength, which prevents it from accepting a hole. Thus, the dynamics are akin to hole motion from Br into the \ce{-C6H4-} ring, and back again. As shown in Fig.~\ref{fig:funct}(b), the \ce{-Si(CH3)3} (trimethylsilyl; TMS) molecule also has qualitatively different dynamics from the other molecules, which involves the beating of at least two frequencies. This can be understood in terms of the very weak electron donating strength of TMS, which makes the molecule act as two decoupled regions. The remaining molecules all behave similarly, and exhibit CM that consists of hole motion from the Br to \ce{-R} group that takes approximately 1~fs. In these plots, CM appears as a spatially and temporally separated hole on Br, followed by a delocalized hole on the ring, then leading to spatiotemporally localized hole on R. Since we do not have nuclear motion or other dephasing effects, the CM oscillates indefinitely. 
Strikingly, as the electron donating strength of the {\it para}-functional group increases, there is a clear increase in the hole density on -R, visible in Figure 1 as increasingly spatiotemporarily localized holes.

Before analyzing the relationship between hole contrast and electron donating strength in detail, we briefly discuss the mechanism by which CM occurs in these systems. Three snapshots of the hole density in 4-bromoaniline are shown in Figure~\ref{fig:mecha}. To emphasize the density changes corresponding to CM, as with the contrast calculations, we slice the data at a distance of 1~\AA\ above the plane. The initial localized hole on Br takes approximately 0.6~fs to move into the phenyl ring, at which time it becomes delocalized across the entire ring. The delocalization across the ring (as opposed to $\pi$-hopping~\cite{Folorunso2021Molecular}) is a consequence of the symmetric shape of the molecule, with the phenyl group containing $\pi$ bonds. After another 0.33~fs the hole then migrates to the opposite end of the molecule, wherein it becomes localized above/below the \ce{NH2} group. This overall time scale is consistent with previously reported CM in benzene~\cite{despre2015attosecond}. A similar mechanism is observed for all the molecules that exhibit CM. The observation that the \ce{-NH2} group supports a local hole at particular times suggests it has a strong hole affinity. 

Next, to quantify the hole affinities for various functional groups, we draw a parallel to the conventional chemical definition of electron withdrawing strength. In substituted benzene rings, each \ce{-R} group can be assigned a Hammett $\sigma$ value, which is a way of quantifying how a particular electron-donating or electron-withdrawing group affects the chemical reactivity of a molecule\cite{hammett1937,Hammettvalue1991}. For an arbitrary reaction, the Hammett equation is $\log\frac{K}{K_0} = \sigma \rho$, where $K$ and $K_0$ are the equilibrium constants (or rate constants) for a molecule functionalized with a particular R group, and a reference functional group, respectively. $\sigma$ is the Hammett substituent constant (one per \ce{R}), and $\rho$ is the reaction-specific constant (one per reaction). To construct a CM analog of $\sigma$, we use a Hammett-like equation, where instead of the chemical reaction rate we use the CM contrast:
\begin{equation}
    \label{eq:cmsigm}
    \sigma^{\Gamma} = \log\frac{\mathrm{\Gamma}}{\mathrm{\Gamma_{\text{0}}}},
\end{equation}
where $\sigma^{\Gamma}$ is the hole constrast sigma value, $\Gamma$ is the hole contrast of specific R group and $\Gamma_0$ is the reference hole contrast(\ce{-Si(CH)3}) . Typically the Hammett $\sigma$ is referenced to benzene ({\it i.e.}, \ce{-R} = \ce{-H})\cite{hammett1937}, but bromobezene does not support CM, and instead involves a hole delocalized across the entire ring. Therefore, we use the hole contrast in 1-Bromo-4-(trimethylsilyl)benzene (R =\ce{-Si(CH)3}) for $\Gamma_0$, since trimethylsilyl (TMS) is predicted to be a very weak hole acceptor. Thus, in our scale, $\sigma^{\Gamma}$ values that are negative have a better hole affinity than TMS, while functional groups with positive values have a worse hole affinity.
\begin{figure}
\centering
  \includegraphics[width=0.85\linewidth]{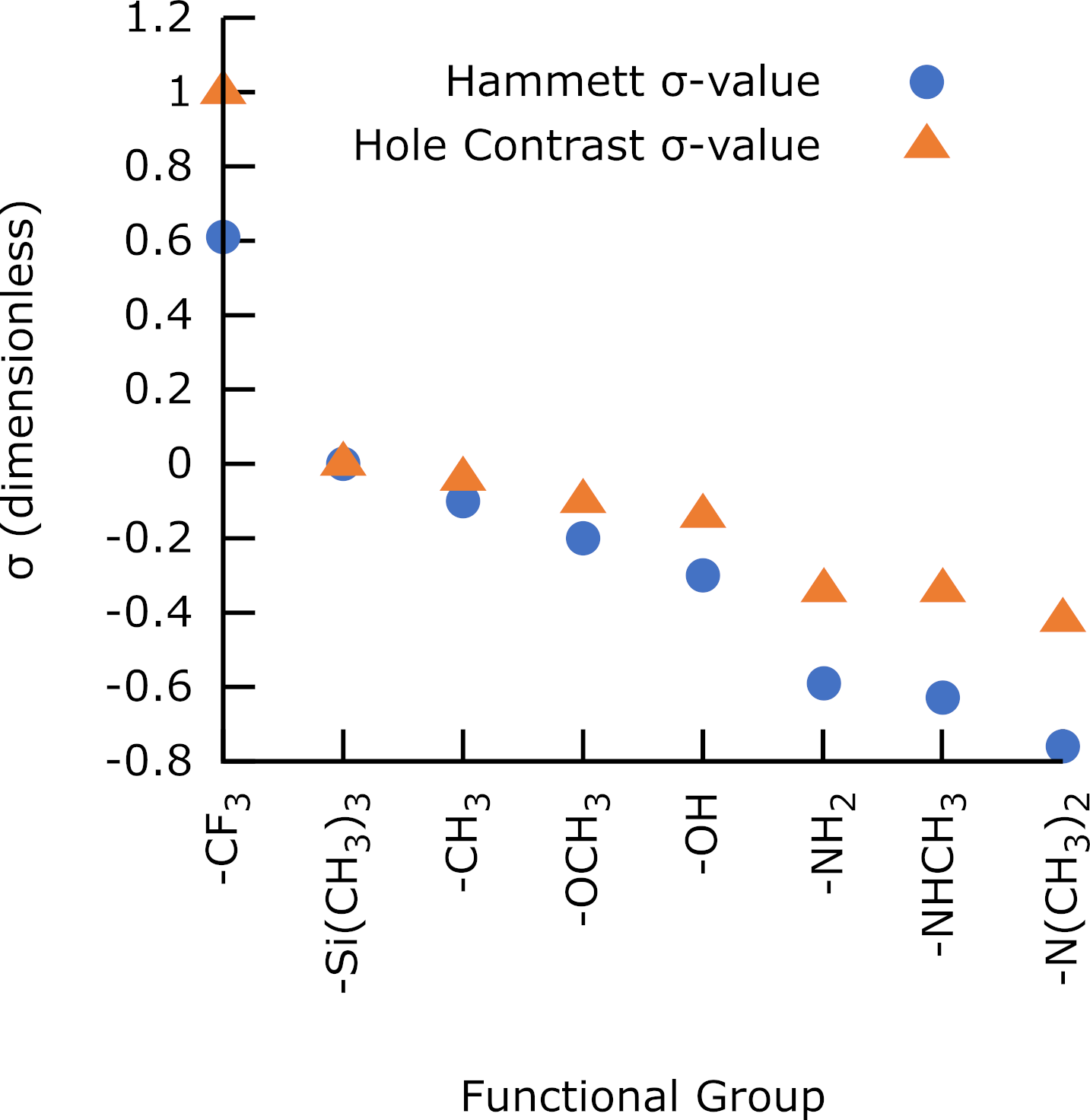}
  \caption{The Hammett and hole contrast sigma values for {\it para} \ce{Br-C6H4-R} for various functional groups. More negative values indicate stronger electron donating strength/hole contrast, respectively. The data has been referenced to \ce{-Si(CH3)3} (see text).}
  \label{fig:hamm}
\end{figure}

Fig.~\ref{fig:hamm} shows the organic chemistry literature Hammett and our computed CM contrast $\sigma$ values, both referenced to TMS. There is good qualitative agreement between the two quantities, with hole contrast $\sigma^{\Gamma}$ decreasing monotonically with decreasing Hammett $\sigma^H$. As visible in the dynamics plots in Fig.~\ref{fig:funct}, \ce{-CF3}, which has a large positive value (strong electron acceptor) is a bad hole acceptor (high $\sigma^{\Gamma}$). \ce{-CH3}, \ce{-OCH3}, \ce{-OH} are all relatively weak electron donors (small negative $\sigma^H$) and thus modest CM hole acceptors. This is a consequence of the high electronegativity of these groups being offset by electron donation via resonance. The amine derivatives \ce{-NH2}, \ce{-NHCH3} and \ce{-N(CH3)2} are highly electron donating (large negative $\sigma^H$) due to the presence of lone pairs which they can easily donate and thus, have a correspondingly good CM hole affinity. Furthermore, adding more methyl groups to the N-atom increases their electron donating strength, because the \ce{-CH3} groups donate electron density  to the nitrogen resulting in increased hole affinity for {\it N}-methylamino and {\it N,N}-dimethylamino relative to amino. This gives a more negative sigma value (good hole acceptor), which is visible as dark red hole densities around the \ce{-R} group in the dynamics plots in Fig.~\ref{fig:funct}(f,g,h). These results are in agreement with previous studies on differential hole mobility in doped conjugated molecules, where n- and p-type doping was observed to modulate hole motion \cite{BNbriged2011}.
It is interesting to note that, at least for the cases presented here, chemical functionalization drastically modifies the hole contrast without significantly affecting the CM time. This makes systematic functionalization a promising avenue for experimental measurements that are sensitive to local electron density at different ends of the molecule ({\it e.g.}, transient X-ray absorption, high harmonic generation, ionization spectroscopy, {\it etc}). On a fundamental level, the surprisingly good correlation between electron withdrawing strength and hole contrast is quite illuminating, as it suggests that simple chemical principles that dictate density distributions in molecules can be good predictors of attosecond electron dynamics, at least for CM which occurs via particle-like motion.

In conclusion, we have used first-principles simulations to determine the effect of chemical functionalization on halogen-centred strong-field ionization triggered CM in {\it para}-functionalized bromobenzene derivatives. In the molecules that do support CM, the observed dynamics involve the movement of the hole across the molecular backbone in a Br localized $\rightarrow$ ring delocalized $\rightarrow$ R localized manner, consistent with previous studies that have shown that CM occurs via a hole propagating in the $\pi$ system of conjugated molecules\cite{Folorunso2021Molecular}. The main observation of this work is that functionalization with groups of varying electron withdrawing strength only slighly modifies the CM speed, but has a pronounced effect on the hole contrast, with strong electron donating groups supporting higher contrast CM.

Our findings have numerous implications. From a practical standpoint, they suggest that hole acceptor functional groups can be used as regulators of CM, and for enhancing observability of the hole, all without changing the CM time scale. In particular, we predict that the family of bromobenzene derivatives with strong electron donating functional groups (especially amines) will be excellent molecules for experimental measurements that seek to probe the local electron density at different ends of the molecule. On the other hand, from an interpretation standpoint the simple qualitative relationship between electron donating strength (Hammet sigma value) and hole contrast bolsters the idea CM can be understood in terms of chemically influenced electron density motion. This density-based ``attochemistry'' picture of CM complements emerging resonance-based~\cite{Folorunso2021Molecular,bruner2017attosecond} (hopping of $\pi$ bonds) and nonlinear multielectron pictures~\cite{mauger2022charge}, both of which describe CM in terms of the electron density alone, without resorting to an ambiguous interpretation in terms of a complicated beating of many states.

\section{Acknowledgments}
This  work  was  supported  by  the  U.S. Department  of  Energy, Office  of Science,  Basic  Energy  Sciences,  under Award  No. DE-SC0012462.  Portions of  this  research were conducted  with  high performance computational resources provided by Louisiana State University (www.hpc.lsu.edu)  and  the Louisiana Optical  Network Infrastructure (www.loni.org).

\bibliography{bibliography.bib}
\end{document}